\documentstyle[aps,epsf]{revtex}  % DO NOT DELETE THIS LINE 
 
\newcommand{\ri}{\mbox{$\rm i$}}

\newcommand{\bfm}[1]{\mbox{\boldmath$#1$}}
\newcommand{\ratio}[2]{\mbox{$#1\over#2$}}

\begin{document}        %  DO NOT DELETE OR CHANGE THIS LINE

\baselineskip 14pt
\title{ 
\hfill \raisebox{2ex}{\rm\small ISU-HET-99-1} \bigskip \\   
$|\Delta\bfm{I}|=3/2$  Decays of Hyperons in Chiral 
Perturbation Theory\thanks{Talk presented at DPF `99, Los Angeles, 
5-9 January 1999.}
}
\author{Jusak Tandean}
\address{Department of Physics and Astronomy, Iowa State University, 
Ames, IA 50011}
% \author{}   % Use this and the next line only if there is a second
% \address{Another University, etc.}  % address. (Remove the left % marks)
%
\maketitle              % Creates the title area, Do Not Remove

\begin{abstract}        % Do Not Delete this line
We study the  $|\Delta\bfm{I}|=3/2$  amplitudes of the  octet-hyperon 
decays  $B\rightarrow B'\pi$  and  of the  decays  
$\Omega^-\rightarrow\Xi\pi$  in the  context of heavy-baryon chiral 
perturbation theory.   
For the  octet-hyperon decays, we investigate the  theoretical 
uncertainty of the  lowest-order predictions by calculating the  
leading nonanalytic corrections. 
We find that these corrections are within the  expectations of naive 
power counting and, therefore, that this picture can be tested more 
accurately with improved measurements.
For the  $\Omega^-$  decays, we obtain at leading order two operators 
responsible for the  decays which also contribute at one loop to the  
octet-hyperon decays.  
These one-loop contributions are sufficiently large to suggest that
the  measured ratio 
$\Gamma(\Omega^-\rightarrow\Xi^0\pi^-)/
\Gamma(\Omega^-\rightarrow\Xi^-\pi^0)\approx 2.7$ 
may be too large.   
\end{abstract}   	% Do Not Delete this line

\section{Introduction}               % Introduction goes below.

Nonleptonic decays of hyperons have been studied by various authors 
in the  framework of chiral perturbation theory~($\chi$PT).  
For the  hyperons belonging to the  baryon octet, the  decay modes are
$\,\Sigma^+\rightarrow n\pi^+,\,$  $\,\Sigma^+\rightarrow p\pi^0,\,$  
$\Sigma^-\rightarrow n\pi^-,\,$  
$\Lambda\rightarrow p\pi^-,\,$  $\Lambda\rightarrow n\pi^0,\,$  
$\Xi^-\rightarrow\Lambda\pi^-,\,$ 
and  $\,\Xi^0\rightarrow\Lambda\pi^0.\,$
Calculations of the  dominant  $\,|\Delta\bfm{I}| =1/2\,$  
amplitudes of these decays have led to mixed 
results~\cite{bijnens,books,jenkins,JenMan2,CarGeo,springer}. 
Specifically, the  theory can give a~good description of either 
the  S-waves or the  P-waves, but not both simultaneously.  
Now, while these amplitudes have been much studied in~$\chi$PT, 
the  same cannot be said of their $\,|\Delta\bfm{I}|=3/2\,$  
counterparts.   
In view of the  situation in the  $\,|\Delta\bfm{I}|=1/2\,$  sector,  
it is instructive to carry out a~similar analysis of 
the  $\,|\Delta\bfm{I}|=3/2\,$  amplitudes. 
Such an analysis has been done recently~\cite{etv}, 
and some of its results will be presented here.

In the  baryon-decuplet sector, only the  $\Omega^-$  hyperon
decays weakly.   
For  $\,\Omega^-\rightarrow\Xi\pi$  decays,  a~purely  
$\,|\Delta\bfm{I}|=1/2\,$  weak interaction would imply  
the  ratio of decay rates  
$\,\Gamma(\Omega^-\rightarrow\Xi^0\pi^-)/ 
\Gamma(\Omega^-\rightarrow\Xi^-\pi^0)=2.\,$ 
Instead, this ratio is measured to be approximately  $2.7\;$~\cite{pdb},  
which seems to suggest that the  $\,|\Delta\bfm{I}|=1/2\,$  rule 
is violated in $\Omega^-$  decays~\cite{CarGeo}. 
This situation has recently been examined in some detail~\cite{TanVal} 
using~$\chi$PT.    
The  result will also be presented here, for the  couplings 
generating the  $\,|\Delta\bfm{I}|=3/2\,$  decays of 
the  $\Omega^-$ also contribute to the  octet-hyperon decays.

To apply~$\chi$PT to interactions involving the  lowest-lying mesons 
and baryons, we employ the  heavy-baryon 
formalism~\cite{JenMan2,JenMan1}.  
In this approach, the  theory has a~consistent chiral expansion, 
and the  octet and decuplet baryons in the  effective chiral Lagrangian  
are described by velocity-dependent fields.   
We include the  decuplet baryons in the  Lagrangian because 
the  octet-decuplet mass difference is small enough to make their 
effects significant on the  low-energy theory~\cite{JenMan2,JenMan3}.

\section{$|\Delta\bfm{I}|=3/2$  Decays of Octet Hyperons}

The  leading-order chiral Lagrangian for the  strong interactions is well 
known~\cite{JenMan2,JenMan1}, and so we will discuss only 
the  weak sector.       
Within the  standard model,  the  $\,|\Delta S|=1$,  
$\,|\Delta\bfm{I}|=3/2\,$  weak transitions are induced by an effective  
Hamiltonian that transforms as  $(27_{\rm L}^{},1_{\rm R}^{})$  
under chiral rotations. 
At lowest order in  $\chi$PT,  the  Lagrangian that describes 
such weak interactions of baryons and has the  required 
transformation properties is~\cite{etv,heval}  
\begin{eqnarray}   
{\cal L}^{\rm w}  \;=\;     
\beta_{27}^{}\, 
T_{ij,kl}^{} \left( \xi\bar{B}_v^{} \xi^\dagger \right) _{\!ki}^{} 
\left( \xi B_v^{} \xi^\dagger \right) _{\!lj}^{}   
\,+\,  
\delta_{27}^{}\, 
T_{ij,kl}^{}\; \xi_{kd}^{} \xi_{bi}^\dagger\; 
\xi_{le}^{} \xi_{cj}^\dagger\; 
\bigl( \bar{T}_v^\mu \bigr) _{abc}^{} 
\bigl( T_{v\mu}^{} \bigr) _{ade}^{} 
\;+\;  {\rm h.c.}   \;, 
\label{loweak}
\end{eqnarray}      
where  $\beta_{27}^{}$  ($\delta_{27}^{}$)  is the  coupling constant
for the  baryon-octet (baryon-decuplet) sector, 
and  $T_{ij,kl}^{}$  is the  tensor that project out  
the  $\,|\Delta S|=1$,  $\,|\Delta\bfm{I}|=3/2\,$  transitions 
(further details are given in Ref.~\cite{etv}).

We now turn to the  calculation of the  amplitudes.  
In the  heavy-baryon approach, the  amplitude for the  decay  
$\,B\rightarrow B^\prime\pi\,$  can be written as~\cite{etv}  
\begin{eqnarray}        
\ri {\cal M}^{}_{B_{}\rightarrow B_{}'\pi}   \;=\;  
G_{\rm F}^{} m_{\pi}^2\, 
\bar{u}_{B_{}'}^{} \left( 
{\cal A}^{(\rm S)}_{B_{}^{}B_{}'\pi}   
+ 2 k\cdot S_v^{}\, {\cal A}^{(\rm P)}_{B_{}^{}B_{}'\pi} 
\right) u_{B_{}^{}}^{}   \;,  
\end{eqnarray}    
where  the  superscripts refer  to S- and P-wave contributions, 
the  $u$'s  are baryon spinors, 
$k$  is the  outgoing four-momentum of the  pion, and  
$S_v^{}$  is the  velocity-dependent spin operator~\cite{JenMan1}.

At tree level,  ${\cal O}(1)$  in  $\chi$PT,  contributions to 
the  amplitudes come from diagrams each with a~weak vertex from 
${\cal L}^{\rm w}$  in~(\ref{loweak}) and, for the  P-waves, 
a vertex from the  lowest-order strong Lagrangian.  
At next order in $\chi$PT,  there are amplitudes of order  $m_s^{}$, 
the  strange-quark mass,  arising both from one-loop diagrams with 
leading-order vertices and from counterterms. 
Currently there is not enough experimental input to determine 
the  value of the  counterterms. 
For this reason, we follow the  approach that has been used for 
the  $\,|\Delta\bfm{I}|=1/2\,$  amplitudes~\cite{bijnens,jenkins} 
and  calculate only nonanalytic terms up to  
${\cal O}(m_s^{}\ln{m_s^{}})$.  
These terms are uniquely determined from the  one-loop amplitudes   
because they cannot arise from local counterterm Lagrangians.   
With a~complete calculation at next-to-leading order, it would be
possible to fit all the  amplitudes 
(as was done in Ref.~\cite{BorHol} for the  $\,|\Delta\bfm{I}|=1/2\,$ 
sector),  but we feel that this exercise is not 
instructive given the  large number of free parameters available. 
In this work, we limit ourselves to study the  question of whether 
the  lowest-order predictions are subject to large higher-order  
corrections.

To compare our theoretical results with experiment, 
we introduce the  amplitudes~\cite{jenkins}
\begin{eqnarray}   
s  \;=\;  {\cal A}^{\rm (S)}   \;, \hspace{2em}   
p  \;=\;  -|\bfm{k}| {\cal A}^{\rm(P)}   \;,      
\end{eqnarray}      
in the  rest frame of the  decaying baryon.    
From these amplitudes, we can extract for the  S-waves 
the  $\,|\Delta\bfm{I}|=3/2\,$  components 
\begin{eqnarray}   
\begin{array}{c}   \displaystyle        
S_{3}^{(\Lambda)}  \;=\;  
\ratio{1}{\sqrt{3}} 
\Bigl( \sqrt{2}\, s_{\Lambda\rightarrow n\pi^0}^{}   
      + s_{\Lambda\rightarrow p\pi^-}^{} \Bigr)   \;,    
\hspace{3em} 
S_{3}^{(\Xi)}  \;=\;  
\ratio{2}{3} \Bigl( \sqrt{2}\, s_{\Xi^0\rightarrow\Lambda\pi^0}^{}  
                   + s_{\Xi^-\rightarrow\Lambda\pi^-}^{} \Bigr)   \;,  
\vspace{2ex} \\   \displaystyle  
S_{3}^{(\Sigma)}  \;=\;  
-\sqrt{\ratio{5}{18}} 
\Bigl( s_{\Sigma^+\rightarrow n\pi^+}^{}  
      - \sqrt{2}\, s_{\Sigma^+\rightarrow p\pi^0}^{}   
      - s_{\Sigma^-\rightarrow n\pi^-}^{} \Bigr)   \;,
\end{array}   
\end{eqnarray}      
and the  $\,|\Delta\bfm{I}|=1/2\,$  components  
(for  $\Lambda$  and  $\Xi$  decays)  
\begin{eqnarray}   
S_{1}^{(\Lambda)}  \;=\;  
\ratio{1}{\sqrt{3}} 
\Bigl( s_{\Lambda\rightarrow n\pi^0}^{}   
      - \sqrt{2}\, s_{\Lambda\rightarrow p\pi^-}^{} \Bigr)   \;,    
\hspace{3em} 
S_{1}^{(\Xi)}  \;=\;  
\ratio{\sqrt{2}}{3} 
\Bigl( s_{\Xi^0\rightarrow\Lambda\pi^0}^{}  
       - \sqrt{2}\, s_{\Xi^-\rightarrow\Lambda\pi^-}^{} \Bigr)   \;, 
\end{eqnarray}      
as well as analogous ones for the  P-waves. 
We can then compute from data the  ratios collected in  
Table~\ref{ratio,isosymmetric}, which show 
the  $\,|\Delta\bfm{I}|=1/2\,$  rule for hyperon decays.  
The  experimental values for  $S_3^{}$  and  $P_3^{}$  
are listed in the  column labeled ``Experiment'' in  
Table~\ref{results1}.

\begin{center}\begin{minipage}{0.9\textwidth}  
\begin{table}
\caption{\label{ratio,isosymmetric}%     
Experimental values of ratios of  $\,|\Delta\bfm{I}|=3/2\,$  to  
$\,|\Delta\bfm{I}|=1/2\,$  amplitudes.
}    
\begin{tabular}{cccccc} 
$S_3^{(\Lambda)}/S_{1_{}}^{(\Lambda)^{}}$            & 
$S_3^{(\Xi)}/S_1^{(\Xi)}$                            & 
$S_3^{(\Sigma)}/s_{\Sigma^-\rightarrow n\pi^-}^{}$   &  
$P_3^{(\Lambda)}/P_1^{(\Lambda)}$                    &   
$P_3^{(\Xi)}/P_1^{(\Xi)}$                            &   
$P_3^{(\Sigma)}/p_{\Sigma^+\rightarrow n\pi^+}^{}$             
\vspace{0.5ex} \\ \hline \vspace{-2ex} \\
$0.026\pm 0.009$   
& $0.042\pm 0.009$   
& $-0.055\pm 0.020$  
& $0.031\pm 0.037$  
& $-0.045\pm 0.047$  
& $-0.059\pm 0.024$  
\end{tabular}   
\end{table}
\end{minipage}\end{center}

To begin discussing our theoretical results, we note that our 
calculation yields no contributions to the  S-wave amplitudes  
$S_3^{(\Lambda)}$  and  $S_3^{(\Xi)}$,  as shown in 
Table~\ref{results1}.  
This only indicates that the  two amplitudes are predicted to be 
smaller than $S_3^{(\Sigma)}$ by about a~factor of three 
because there are nonvanishing contributions from operators that 
occur at the  next order,  ${\cal O}(m_s^{}/\Lambda_{\chi\rm SB}^{})$, 
with  $\,\Lambda_{\chi\rm SB}^{}\sim 1\,\rm GeV\,$  
being the  scale of chiral-symmetry breaking.
(An example of such operators is considered in Refs.~\cite{etv,heval}.)  
The  experimental values of $S_3^{(\Lambda)}$  and 
$S_3^{(\Xi)}$  are seen to support this prediction.

\begin{center}\begin{minipage}{0.9\textwidth}  
\begin{table}  
\caption{\label{results1}%     
Summary of results for  $\,|\Delta\bfm{I}|=3/2\,$  components of 
the  S- and P-wave  amplitudes to  ${\cal O}(m_s^{}\ln{m_s^{}})$.   
We use the  parameter values 
$\,\beta_{27}^{}=\delta_{27}^{}=-0.068 \, 
\sqrt{2}\, f_{\!\pi}^{}G_{\rm F}^{} m_{\pi}^2\,$
and a~subtraction scale  $\,\mu = 1\,\rm GeV.\,$}  
\begin{tabular}{crrrr}    
&& \multicolumn{3}{c}{Theory}  
\\ \cline{3-5}    
Amplitude  &  Experiment\hspace{1ex}    &  
\raisebox{-1ex}{Tree\hspace{1ex}}       &  
\raisebox{-1ex}{Octet\hspace{1.5em}}    &   
\raisebox{-1ex}{Decuplet\hspace{2ex}}   
\\   
&&  
\raisebox{-1ex}{${\cal O}(1)$\hspace{0.5ex}}       &   
\raisebox{-1ex}{${\cal O}(m_s^{}\ln{m_s^{}})$}     & 
\raisebox{-1ex}{${\cal O}(m_s^{}\ln{m_s^{}})$}        
\smallskip \\ \hline      
\vspace{-2.5ex} \\  
$S_3^{(\Lambda)}$ &  $-0.047\pm 0.017$    &   0\hspace{1em} &   
\hspace{5ex} 0\hspace{2.5em}   &  
\hspace{4ex} 0\hspace{2.5em}   
\\
$S_3^{(\Xi)}$          & $ 0.088\pm 0.020$      & 0\hspace{1em}    &  
0\hspace{2.5em}        & 0\hspace{2.5em}        
\\   
$S_3^{(\Sigma)}$       & $-0.107\pm 0.038$      & $-$0.107         &  
$-$0.089\hspace{1.5em} & $-$0.084\hspace{1.5em} 
\\  
$P_3^{(\Lambda)}$      & $-0.021\pm 0.025$      & 0.012            &  
0.005\hspace{1.5em}    & $-$0.060\hspace{1.5em} 
\\  
$P_3^{(\Xi)}$          & $ 0.022\pm 0.023$      & $-$0.037         &   
$-$0.024\hspace{1.5em} & 0.065\hspace{1.5em}    
\\  
\vspace{.3ex}   
$P_3^{(\Sigma)}$       & $-0.110\pm 0.045$      & 0.032            &   
0.015\hspace{1.5em}    & $-$0.171\hspace{1.5em} 
\\  
\end{tabular}   
\end{table}
\end{minipage}\end{center}

The  other four amplitudes are predicted to be nonzero.   
They depend on the  two weak parameters  $\beta_{27}^{}$  and  
$\delta_{27}$  of  ${\cal L}^{\rm w}\,$  
(as well as on parameters from the  strong 
Lagrangian, which are already determined),  
with  $\delta_{27}$  appearing only in loop diagrams.
Since we consider only the  nonanalytic part of the  loop diagrams, 
and since the  errors in the  measurements of the  P-wave amplitudes 
are larger than those in the  S-wave amplitudes, we can take the  
point of view that we will extract the  value of $\beta_{27}^{}$
by fitting the  tree-level $S_3^{(\Sigma)}$ amplitude to experiment, 
and then treat the  tree-level P-waves as predictions and the  loop 
results as a~measure of the  uncertainties of the  lowest-order 
predictions.

Thus, we obtain 
$\,\beta_{27}^{}=-0.068\,\sqrt{2}\, f_{\!\pi}^{}  
   G_{\rm F}^{} m_{\pi}^2,\,$  
and the  resulting P-wave amplitudes are placed in the  column labeled
``Tree'' in  Table~\ref{results1}.  
These lowest-order predictions are not impressive, but they 
have the  right order of magnitude and differ from the  central 
value of the  measurements by at most three standard deviations.  
For comparison, in the  $\,|\Delta\bfm{I}|=1/2\,$  case   
the  tree-level predictions for the  P-wave amplitudes are completely 
wrong~\cite{bijnens,jenkins}, differing from the  measurements by 
factors of up to 30.

To address the  reliability of the  leading-order predictions, 
we look at our calculation of the  one-loop corrections,    
presented in two columns in  Table~\ref{results1}.  
The  numbers in the  column marked  ``Octet''  come from  
all loop diagrams that do not have any decuplet-baryon lines,  
with  $\beta_{27}^{}$  being the  only weak parameter in the  diagrams.  
Contributions of loop diagrams with decuplet baryons depend on one 
additional constant, $\delta_{27}^{}$,  which cannot be fixed 
from experiment as it does not appear in any of the  observed 
weak decays of a~decuplet baryon. 
To illustrate the  effect of these terms, we choose  
$\,\delta_{27}^{}=\beta_{27}^{},\,$  a~choice consistent with 
dimensional analysis and the  normalization of  ${\cal L}^{\rm w}$,  
and collect the  results in the  column labeled ``Decuplet''.

We can see that some of the  loop corrections in  Table~\ref{results1}  
are comparable to or even larger than the  lowest-order results 
even though they are expected to be smaller by about a~factor of 
$\,M_K^2/(4\pi f_{\!\pi}^{})^2\approx 0.2.\,$   
These large corrections occur when several different diagrams 
yield contributions that add up constructively, resulting in 
deviations of up to an order of magnitude from the  power-counting 
expectation. 
This is an inherent flaw in a~perturbative calculation where 
the  expansion parameter is not sufficiently small.      
We can, therefore, say that these numbers are consistent 
with naive expectations.

Although the  one-loop corrections are large, they are all much 
smaller than their counterparts in  $\,|\Delta\bfm{I}|=1/2\,$  
transitions, where they can be as large as 15 times 
the  lowest-order amplitude in the  case of the  P-wave in  
$\,\Sigma^+ \rightarrow n \pi^+.\,$ 
In that case, the  discrepancy was due to an anomalously 
small lowest-order prediction arising from the  cancellation of two 
nearly identical terms~\cite{jenkins}.

In conclusion, we have presented a~discussion of  
$\,|\Delta\bfm{I}|=3/2\,$  amplitudes for hyperon nonleptonic 
decays in $\chi$PT. 
At leading order these amplitudes are described in terms of 
only one weak parameter. 
This parameter can be fixed from the  observed value of 
the  S-wave amplitudes in $\Sigma$ decays.  
After fitting this number, we have predicted the  P-waves and 
used our one-loop calculation to discuss uncertainties of 
the  lowest-order predictions. 
Our predictions are not contradicted by current data, but current 
experimental errors are too large for a~meaningful conclusion.   
We have shown that the  one-loop nonanalytic corrections have 
the  relative size expected from naive power counting. 
The~combined efforts of E871 and KTeV experiments at Fermilab could 
give us improved accuracy in the  measurements of some of the  decay 
modes that we have discussed and allow a~more quantitative 
comparison of theory and experiment.

\section{$|\Delta\bfm{I}|=3/2$  Decays of the  $\Omega^-$}

In the  heavy-baryon formalism, we can write the  amplitude for  
$\,\Omega^-\rightarrow\Xi\pi\,$  as~\cite{TanVal}
\begin{eqnarray}   \label{amplitude}     
\ri {\cal M}_{\Omega^-\rightarrow\Xi\pi}^{}  \;=\;  
G_{\rm F}^{} m_{\pi}^2\; \bar{u}_\Xi^{}\, 
{\cal A}_{\Omega^-\Xi\pi}^{\rm (P)}\, k_\mu^{}\, u_\Omega^\mu   
\;\equiv\;  
G_{\rm F}^{} m_{\pi}^2\; \bar{u}_\Xi^{}\, 
{\alpha_{\Omega^-\Xi}^{\rm (P)}\over \sqrt{2}\, f_{\!\pi}^{}}\,  
k_\mu^{}\, u_\Omega^\mu      \;,
\end{eqnarray}    
where  the  $u$'s  are baryon spinors,  $k$  is the  outgoing four-momentum 
of the  pion,  and  only the  dominant P-wave piece of the  amplitude 
is included.  
We will consider only the  P-wave because, experimentally, 
the  asymmetry parameter in these decays is small and consistent with 
zero~\cite{pdb}, indicating that they are dominated by a~P-wave.

From the  measured decay rates,  we obtain~\cite{TanVal}  
\begin{eqnarray}         
{\cal A}^{({\rm P})}_{\Omega^-\Xi^-\pi^0}  \;=\;   
(3.31\pm 0.08) \;{\rm GeV}^{-1}  
\;, \hspace{3em}  
{\cal A}^{({\rm P})}_{\Omega^-\Xi^0\pi^-}  \;=\;   
(5.48\pm 0.09) \;{\rm GeV}^{-1}   \;.     
\end{eqnarray}   
Upon defining the  $\,|\Delta\bfm{I}|=1/2,\, 3/2\,$  amplitudes 
\begin{eqnarray}   \label{a1,a3}      
\alpha^{(\Omega)}_1  \;\equiv\; 
\ratio{1}{\sqrt{3}} 
\left( \alpha^{({\rm P})}_{\Omega^-\Xi^-} 
      + \sqrt{2}\, \alpha^{({\rm P})}_{\Omega^-\Xi^0} \right)   
\;, \hspace{3em}  
\alpha^{(\Omega)}_3  \;\equiv\; 
\ratio{1}{\sqrt{3}} 
\left( \sqrt{2}\, \alpha^{({\rm P})}_{\Omega^-\Xi^-} 
      - \alpha^{({\rm P})}_{\Omega^-\Xi^0} \right)   \;,   
\end{eqnarray}   
respectively, we can extract the  ratio
\begin{eqnarray}   \label{ratio,o}
\alpha_{3}^{(\Omega)}/\alpha_{1}^{(\Omega)}  \;=\;  -0.072\pm 0.013   \;,
\end{eqnarray}      
which is higher than the  corresponding ratios in octet-hyperon 
decays listed in Table~\ref{ratio,isosymmetric}, 
but not significantly so.

Although the  size of this ratio is not clear evidence for 
violation of the  $\,|\Delta\bfm{I}|=1/2\,$  rule in  $\Omega^-$ 
decays, it leads to a~different question, 
that of the  compatibility of the  measurements of these decays and 
those of the  octet-hyperon decays.   
To address this question, we will first extract 
a $\,|\Delta\bfm{I}|=3/2\,$  coupling from 
$\,\Omega^-\rightarrow\Xi\pi\,$  decays and then examine its 
contribution to the  octet-hyperon decays.

Employing standard group-theory techniques, we find two different 
operators that transform as  $(27_{\rm L}^{},1_{\rm R}^{})$  
and generate  $\,\Delta S=1,$  $\,|\Delta\bfm{I}|=3/2\,$  
transitions involving  $\Omega^-$  fields.  
We write them as 
\begin{eqnarray}   \label{l1weak} 
{\cal L}_1^{\rm w}  \;=\;   
T_{ij,kl}^{}\; \xi_{ka}^{} \xi_{lb}^{} 
\left( {\cal C}_{27}^{}\, I_{ab,cd}^{} + {\cal C}_{27}'\, I_{ab,cd}' \right)   
\xi_{ci}^\dagger \xi_{dj}^\dagger \;,      
\end{eqnarray}      
where  ${\cal C}_{27}^{}$  and  ${\cal C}_{27}'$  are the  weak   
parameters for the  two operators, the  baryon fields are contained 
in the  tensors~$I$  and~$I'$,  and  additional details can be 
found in Ref.~\cite{TanVal}.   
This Lagrangian contains the  terms
\begin{eqnarray}    
{\cal L}_{\Omega^- B\phi}^{\rm w}  &=&       
{{\cal C}_{27}^{}\over f}\, 6
\left( -\sqrt{2}\, \bar{\Sigma}_v^-\, \partial^\mu K^0 
      + 2\, \bar{\Sigma}_v^0\, \partial^\mu K^+ 
      - 2\, \bar{\Xi}_v^-\, \partial^\mu \pi^0 
      + \sqrt{2}\, \bar{\Xi}_v^0\, \partial^\mu \pi^+ \right) \Omega_{v\mu}^-
\nonumber \\ && 
+\;  
{{\cal C}_{27}'\over f}\, 2
\left( \sqrt{2}\, \bar{\Sigma}_v^-\, \partial^\mu K^0 
      - 2\, \bar{\Sigma}_v^0\, \partial^\mu K^+ 
      - 2\, \bar{\Xi}_v^-\, \partial^\mu \pi^0 
      + \sqrt{2}\,\bar{\Xi}_v^0\,\partial^\mu\pi^+ \right) \Omega_{v\mu}^-   \;. 
\end{eqnarray}      
From this expression, one can see that the  decay modes  
$\,\Omega^-\rightarrow\Xi\pi\,$  measure the  combination  
$\,3{\cal C}_{27}+{\cal C}^\prime_{27}.\,$   
Since the  decays  $\,\Omega^-\rightarrow\Sigma K\,$  are
kinematically forbidden, and since three body decays of the  $\Omega^-$
are poorly measured, it is not possible at present to extract 
these two constants separately.

At tree level, the  P-wave amplitudes arise from   
contact diagrams generated by  ${\cal L}_1^{\rm w}$  in  
(\ref{l1weak})  and are given by  
\begin{eqnarray}   \label{contact}      
\alpha^{({\rm P})}_{\Omega^-\Xi^-}  \;=\;   
-4\sqrt{2} \left( 3{\cal C}_{27}^{} + {\cal C}_{27}' \right)  
\;, \hspace{3em}  
\alpha^{({\rm P})}_{\Omega^-\Xi^0}  \;=\;   
4 \left( 3{\cal C}_{27}^{} + {\cal C}_{27}' \right)   \;.  
\end{eqnarray}   
The  value of  the  constant  $\,3{\cal C}_{27}^{}+{\cal C}_{27}'\,$   
is then found to be  
\begin{eqnarray}         
3{\cal C}_{27}^{} + {\cal C}_{27}'  \;=\;  
(8.7 \pm 1.6)\times 10^{-3}\; G_{\rm F}^{} m_{\pi}^2   \;.    
\label{fitom}
\end{eqnarray}   
This value is consistent with power counting, being suppressed by 
approximately a~factor of  $\Lambda_{\chi\rm SB}^{}$ with respect to 
the  parameter  $\beta_{27}^{}$  previously discussed.

We now address the  question of the  size of the  contribution of  
${\cal L}_1^{\rm w}$  in~(\ref{l1weak})  to 
the  $\,|\Delta\bfm{I}|=3/2\,$  decays of octet hyperons at one-loop.
We again keep only the  nonanalytic terms of the  loop results.  
As an illustration of the  effect of these terms on the  octet-hyperon 
decays,  we present numerical results in Table~\ref{result1}, 
where  we look at four simple scenarios to 
satisfy  Eq.~(\ref{fitom})  in terms of only one parameter. 
Interestingly, there are no contributions to  
$S_3^{(\Lambda)}$  and  $S_3^{(\Xi)}$  as before,  and so 
only the  amplitudes predicted to be nonzero are displayed.  
For comparison, we show in the  same Table the  experimental value of 
the  amplitudes as well as the  best theoretical fit at  
${\cal O}(m_s\log m_s)$  obtained in Ref.~\cite{etv}.

\begin{center}\begin{minipage}{0.9\textwidth}  
\begin{table}[b]      
\caption{\label{result1}%  
New $\,|\Delta\bfm{I}|=3/2\,$  contributions to 
S- and P-wave hyperon decay  amplitudes compared with experiment and
with the  best theoretical fit of Ref.~\protect\cite{etv}.  
Here  ${\cal C}_{27}^{}$  and  ${\cal C}_{27}'$  are given in units 
of  $\,10^{-3}\;G_{\rm F}^{} m_{\pi}^2,\,$  and their values are chosen 
to fit the  $\,\Omega^-\rightarrow\Xi\pi\,$  decays.
}  
\begin{tabular}{rcrrrrrrrrrrrr}    
\vspace{-2ex} \\  
&&& Theory\hspace{-1ex} && 
\multicolumn{9}{c}{Theory, new contributions with  
$\,3{\cal C}_{27}^{}+{\cal C}_{27}'=8.7\,$} 
\vspace{1ex} \\ \cline{6-14}    
& 
\raisebox{2ex}[2ex]{Amplitude}  &  
\raisebox{2ex}[2ex]{Experiment\hspace{1ex}}  &  
\raisebox{0.5ex}{Ref.~\cite{etv}\hspace{-1ex}} &&&  
\multicolumn{2}{c}{${\cal C}_{27}'=0$}\hspace{3ex} &
\multicolumn{2}{c}{${\cal C}_{27}^{}=0$}\hspace{3ex} & 
${\cal C}_{27}'={\cal C}_{27}^{}$\hspace{-2.5ex} &&  
\hspace{2ex}${\cal C}_{27}'=-{\cal C}_{27}^{}$\hspace{-4ex} & 
\vspace{0.3ex} \\ \hline      
\vspace{-2.5ex} \\  
& 
$S_3^{(\Sigma)}$   &   $-0.107\pm 0.038$  & $-$0.120  &&&  $-$0.29   &&  
0.52               &&  $-$0.09   && $-$0.70 &         
\\  
& 
$P_3^{(\Lambda)}$  &   $-0.021\pm 0.025$  & $-$0.023  &&& 0.02       &&  
$-$0.04            &&  0.01      && 0.05  &      
\\  
&  
$P_3^{(\Xi)}$      &   $ 0.022\pm 0.023$  & 0.027     &&& $-$0.10    &&   
0.10               &&  $-$0.05   && $-$0.20 &           
\\  
\vspace{.1ex}   
&  
$P_3^{(\Sigma)}$   &   $-0.110\pm 0.045$  & $-$0.066  &&& 0.05       &&   
$-$0.09            &&  0.02      && 0.13  & \hspace*{1em}      
\\  
\end{tabular}   
\end{table}
\end{minipage}\end{center}  

The  new terms calculated here (with  $\,\mu=1\;\rm GeV$), induced by  
${\cal L}_1^{\rm w}$,  are of higher order in  
$m_s^{}$  and  are therefore expected to be at most comparable to 
the  best theoretical fit.  
A quick glance at  Table~\ref{result1}  shows that in some cases 
the  new contributions are much larger.  
Another way to gauge the  size of the  new contributions is to compare 
them  with the  experimental error in the  octet-hyperon decay 
amplitudes.  
Since the  theory provides a~good fit at  
${\cal O}(m_s\log m_s)$~\cite{etv}, we would like the  new 
contributions (which are of higher order in  $m_s^{}$)  to be at most 
at the  level of the  experimental error.  
From  Table~\ref{result1},  we see that in some cases the  new 
contributions are significantly larger than these errors. 
In a~few cases they are significantly larger than the  experimental 
amplitudes.  
All this indicates to us that the  measured  
$\,\Omega^-\rightarrow\Xi\pi\,$  decay rates imply 
a  $\,|\Delta\bfm{I}|=3/2\,$  amplitude that may be too large and 
in contradiction with the  $\,|\Delta\bfm{I}|=3/2\,$  
amplitudes measured in octet-hyperon decays.

Nevertheless, it is premature to conclude that the  measured values for
the  $\,\Omega^-\rightarrow\Xi\pi\,$  decay rates must be incorrect because,
strictly speaking, none of the  contributions to octet-baryon decay
amplitudes is proportional to the  same combination of parameters
measured in  $\,\Omega^-\rightarrow\Xi\pi\,$  decays,  
$\,3 {\cal C}_{27}^{}+{\cal C}_{27}'.\,$  
It is possible to construct linear combinations of the  four amplitudes  
$S_3^{(\Sigma)}$,  $P_3^{(\Sigma)}$,  $P_3^{(\Lambda)}$  and  $P_3^{(\Xi)}$  
that are proportional to  $\,3 {\cal C}_{27}^{}+{\cal C}_{27}'.\,$
We find that the  most sensitive one is   
\begin{equation}
\left( S_3^{(\Sigma)}-4.2 P_3^{(\Xi)} \right) _{\rm Exp}^{}   \;=\;  
-0.2 \pm 0.1   \;,  
\end{equation}
where we have simply combined the  errors in quadrature.  
The  contribution from  ${\cal L}_1^{\rm w}$  to this combination is 
\begin{equation}
\left( S_3^{(\Sigma)}-4.2 P_3^{(\Xi)} \right) _{\rm Theory,new}^{}  
\;\approx\; 
13\, \bigl( 3 {\cal C}_{27}^{}+{\cal C}_{27}' \bigr)  
\;\approx\;  0.1   \;,
\end{equation}  
which falls within the  error in the  measurement.

Our conclusion is that the  current measurement of the  rates for 
$\,\Omega^-\rightarrow\Xi\pi\,$  implies a~$\,|\Delta\bfm{I}|=3/2\,$ 
amplitude that appears large enough to be in conflict with 
measurements of  $\,|\Delta\bfm{I}|=3/2\,$  amplitudes in 
octet-hyperon decays. 
However, within current errors and without any additional 
assumptions about the  relative size of  ${\cal C}_{27}^{}$  and  
${\cal C}_{27}'$,  the~two sets of measurements are not in conflict.

\acknowledgments  
The  material presented here has been drawn from recent papers 
done in collaboration with A.~A.~El-Hady and G.~Valencia.   
This work  was supported in part by DOE under contract number 
DE-FG02-92ER40730.

\end{document}